# SPATIAL INTERPOLATION OF HIGH-FREQUENCY MONITORING DATA[1]

By Michael L. Stein

*University of Chicago*

Climate modelers generally require meteorological information on regular grids, but monitoring stations are, in practice, sited irregularly. Thus, there is a need to produce public data records that interpolate available data to a high density grid, which can then be used to generate meteorological maps at a broad range of spatial and temporal scales. In addition to point predictions, quantifications of uncertainty are also needed. One way to accomplish this is to provide multiple simulations of the relevant meteorological quantities conditional on the observed data taking into account the various uncertainties in predicting a space-time process at locations with no monitoring data. Using a high-quality dataset of minute-by-minute measurements of atmospheric pressure in north-central Oklahoma, this work describes a statistical approach to carrying out these conditional simulations. Based on observations at 11 stations, conditional simulations were produced at two other sites with monitoring stations. The resulting point predictions are very accurate and the multiple simulations produce well-calibrated prediction uncertainties for temporal changes in atmospheric pressure but are substantially overconservative for the uncertainties in the predictions of (undifferenced) pressure.

**1. Introduction.** The US Department of Energy established the Atmospheric Radiation Measurement (ARM) Program to evaluate and improve models for clouds and radiative processes, which are critical components of climate models. The first such site (there are now three) was the Southern Great Plains site established in 1992 in north-central Oklahoma (see www.arm.gov/sites/sgp.stm). Among the many meteorological measurement systems that make up this program is the Surface Meteorological Observation System (SMOS), which records surface wind speed and direction,

Received April 2008; revised June 2008.

[1]Supported wholly or in part by the United States Department of Energy, Office of Science, Office of Biological and Environmental Research, Climate Change Research Division, under contract DE-AC02-06CH11357, as a part of the SciDAC program.

temperature, relative humidity and pressure every minute at a network that currently consists of 23 facilities. However, meteorological modelers are generally more interested in averages of these quantities over grid cells and over longer time scales than every minute. Thus, it is important to develop methods for interpolating the available observations to these spatial and temporal scales. When using such interpolations for model evaluation, it is helpful to have realistic assessments of uncertainty in addition to point predictions. This work represents a small first step of a much larger project (see www.atmos.anl.gov/DMCP/) to provide a publicly available system for generating such predictions and their attendant uncertainties. One approach to doing this is to provide a meteorological equivalent to multiple imputations for censuses with missing data [Rubin (1987)], although we think the name "data ensembles" is more apt than multiple imputations in the present context. Indeed, a recent editorial in the *Bulletin of the American Meteorological Society* [Schneider (2006)] calls for exactly such an approach to providing more useful meteorological data products to the scientific community. Since it is difficult to anticipate all the spatial and temporal scales that might be of interest, a public use dataset should include ensembles of meteorological fields on fine temporal and spatial scales, which could then be aggregated to obtain such fields on a variety of coarser scales. It will be essential for the conditional simulations to capture the dependencies in space–time of the interpolation errors in order to obtain realistic uncertainties for these predictions of aggregated quantities.

This paper considers a very limited effort to produce and evaluate such data ensembles. Rather than producing an ensemble of all the meteorological quantities measured by SMOS, I only consider atmospheric pressure, which avoids the general problem of multivariate spatial-temporal modeling and some of the specific problems of modeling surface winds, which can have erratic patterns in space and time. In addition, I only produce the ensembles for a single month, October 2005, thus limiting the size of the problem and avoiding issues of seasonality. Finally, rather than predicting area averages over some highly resolved set of grid cells, I left out two SMOS sites from the analysis and then predicted pressure at these two sites to evaluate directly the quality of the data ensembles. The data at these two sites were not used in any way whatsoever until after the data ensembles were produced, so comparisons of the resulting data ensembles to the actual observations, favorable or not, provide a fair test of the method's ability to predict pressure over various time scales at unmonitored sites.

This work uses a purely statistical approach with only a minimum of meteorological input (e.g., pressure depends on altitude). Such an approach would be silly for forecasting more than a few hours into the future, but may be difficult to improve on for spatial interpolation in the past, especially in the ARM SGP region for which there is so much data. One might try to improve on this empirical approach by using the pressure given in, for example, the North American Regional Reanalysis (NARR) at NCAR (dss.ucar.edu/pub/narr), which incorporates meteorological measurements into a weather model to produce publicly available records of "hindcasts" of various meteorological quantities. However, the temporal resolution of these records is every 3 hours and the spatial resolution is for grid cells of 32 km $\times$ 32 km, which is coarser in both space and time than we are seeking here. It is, in principle, possible to do a higher resolution version of a hindcast over limited regions such as the ARM SGP domain and it would be interesting to see how much a high-resolution hindcast might improve the kinds of predictions obtained here by, for example, just using these hindcasts as a mean field for pressure. Publicly available hindcasts such as NARR do not provide any direct information about uncertainties in their outputs, so some kind of statistical modeling would still be needed to produce uncertainties, especially at finer spatial resolutions than the spatial grid of the model.

Section 2 discusses some preliminary analyses, such as treatment of missing observations, adjustments for elevation, removal of the diurnal cycle and the changing volatility of pressure in order to obtain a processed form of the data that can be approximated by a stationary Gaussian process in space–time. Section 3 describes the specific form of the Gaussian process model used here, which is an adaptation of a model introduced in Stein (2005). Section 4 presents the results of the data ensembles and shows that they provide very accurate point predictions of pressure at the two sites withheld from the data analysis. The uncertainties across the ensemble members do a good job of mimicking the actual uncertainties of temporal differences in pressure, but have substantially greater variability than the actual prediction errors for undifferenced pressure, although I will argue that this overconservativeness is not necessarily a sign of a problem with the analysis. Section 5 discusses some of the challenges that need to be addressed to produce data ensembles of multiple meteorological quantities at high spatial and temporal resolution.

**2. Preliminary analyses.** Figure 1 shows the locations and elevations of the monitoring sites used in this study. At each of these 13 sites, atmospheric pressure was measured every minute during October 2005, with no more than 8 missing observations in any of the series. I will only use the first 30 days of this month to obtain a highly composite series length of 8640, which speeds up the spectral analyses. Given the tiny fraction of missing observations and the strong continuity in time of the measurements, missing observations were filled in separately at each site using linear interpolation between the nearest available observations before and after each missing observation.

If we could use a Gaussian process model that is stationary in space-time, the inferential and computational problems in obtaining the data ensembles would be greatly simplified. However, it is not appropriate to use such a model directly for the atmospheric pressure process considered here for a number of reasons. First, pressure is recorded to the nearest hundredth of a kilopascal (kPa) and, due to the smoothness in time of pressure and the high precision of the instruments, there is noticeable discreteness in the observations, with the first differences of observed pressure equaling 0 more than 70% of the time. An easy fix for the future would be to record pressure to another significant digit, although one should keep in mind that the overall uncertainty (including various sources of bias) in the measurements has been determined to be $\pm 0.035$ kPa [Ritsche (2008)]. Thus, an extra digit would help to make pressure changes more nearly Gaussian, but it would not help with determining absolute pressure levels. Here, I will consider 5-minute averages of pressure, which, in addition to reducing the discreteness of the data (about 20% of first differences of these are exactly 0), shrinks the dataset by a factor of 5 while maintaining high temporal frequency. Write $Z(\mathbf{x}, t)$ for the 5-minute average of atmospheric pressure at site $\mathbf{x}$ with a time step of 1 corresponding to 5 minutes.

One nonstationary aspect of the data is that mean pressure varies with site. Nearly all of this variation can be explained by variations in altitude;

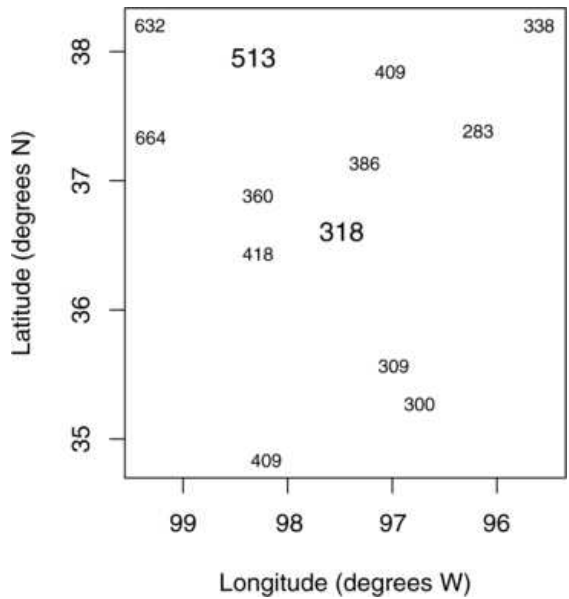

Fig. 1. *Locations of monitoring sites with numbers indicating elevation (m) and large font indicating prediction sites; the prediction site with elevation 318 m will be called the "central" site and the one with elevation 513 m the "peripheral" site.*

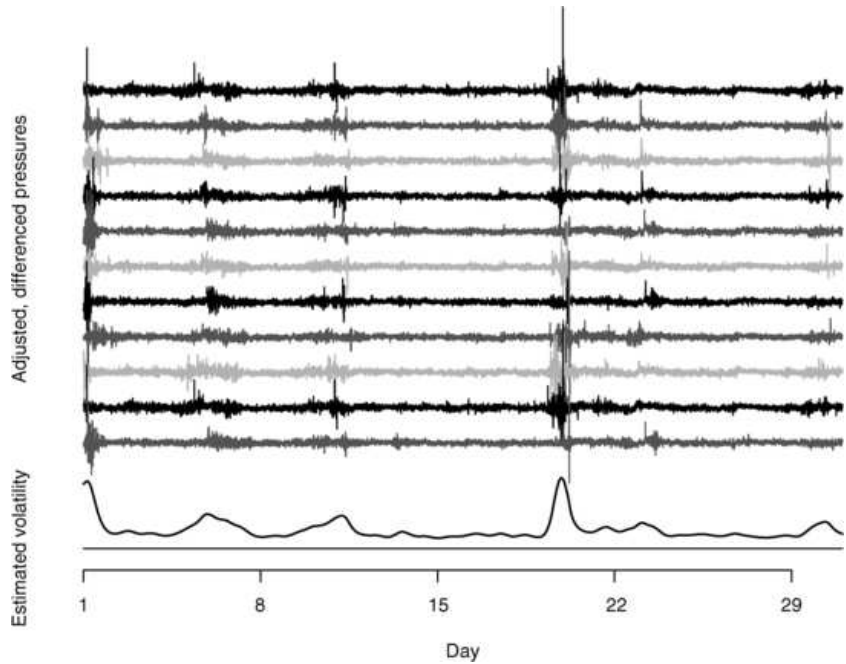

Fig. 2. *First differences of 5-minute averages of elevation-adjusted atmospheric pressure, diurnal cycle removed, for October 2005. Sites, beginning at the top, are arranged from westernmost to easternmost. Curve at the bottom is the estimated volatility function.*

pressure should generally decrease exponentially as altitude increases. Meteorologists commonly use temperature-dependent corrections of surface pressure to sea level pressure, but there is no clear consensus on how best to do this [see, e.g., Mass, Steenburgh and Schultz (1991)], so I will use a simple approach using just altitude here, which appears to work quite well for this single month. Denote by $\bar{Z}(\mathbf{x})$ the average October pressure at $\mathbf{x}$. Regressing the logarithms of these averages on the altitudes of the 11 stations by least squares yields an $R^2$ of 0.9995 and an estimated mean pressure level of $101.89\exp(-a/8310)$ kPa, with $a$ being the altitude of a site in meters. The residuals from this regression show a weak but perceptible spatial pattern and how we handle this pattern will turn out to have a nontrivial impact on our predictions. For now, though, I will focus on modeling the first differences in time of the corrected to sea level 5-minute average pressure, denoted by $D(\mathbf{x}, t)$, on which $\bar{Z}(\mathbf{x})$ has no effect. Differencing pressure may make meteorological interpretation of the statistical model more difficult, but I will argue at the end of this section that it is preferable to modeling undifferenced pressure directly.

Another aspect of the data that must be taken into account is the diurnal cycle. Although not as strong as the diurnal cycle in temperature or relative humidity, it is still quite noticeable in plots of the data. I regressed the average of $D(\mathbf{x},t)$ over the 11 sites on $\cos(2\pi jt/288)$ and $\sin(2\pi jt/288)$ for $j=1,\ldots,15$, with 15 chosen based on numerical and visual inspections. This regression removed an average of 12.4% of the variation in these 11 series. Denote by $R(\mathbf{x},t)$ the residuals from this regression, which are plotted in Figure 2.

It is obvious that these data cannot be plausibly modeled as a stationary Gaussian process due to the occasional bursts of increased variability that occur at least roughly simultaneously at all of the sites. The times with higher volatility are largely related to the passage of weather fronts through the region. To predict future pressure, we would need to model this volatility process, for which it would be crucial to use larger-scale meteorological information. However, since here I only predict at times for which there are observations, I will attempt to remove this volatility empirically by dividing $D(\mathbf{x},t)$ by a function of just time, denoted by $V(t)$. I obtain $V$ by, at each time $t$, computing the sample standard deviation of the 11 available $R(\mathbf{x},t)$ values and then smoothing the logarithms of these standard deviations using a cubic smoothing spline (the R program smooth.spline with the degrees of freedom set to 72). This estimated volatility function is plotted in Figure 2 and, at least qualitatively, it appears to do a good job of tracking the changing volatility of the time series. The use of the logarithmic scale was in part to penalize strongly any very small values for $V(t)$, which could result in small fluctuations in pressure having a large impact on the likelihood. The ratio of the maximum to the minimum value of $V(t)$ is 6.86, so the range of estimated volatilities is quite large. Note that it is not obvious that using *spatial* variability in $R(\mathbf{x},t)$ values will correct for the temporal variability of spread within each series, but this does appear to be largely the case. Although taking $V$ independent of $\mathbf{x}$ throughout this region may be a decent approximation here, in a larger region it would become untenable. However, allowing $V$ to depend on $\mathbf{x}$ would greatly complicate the modeling, especially when one wants to predict pressure at unmonitored locations. Denote by $A(\mathbf{x},t) = R(\mathbf{x},t)/V(t)$ the adjusted residuals. It is these adjusted residuals that I model by a stationary Gaussian process.

Figure 3 shows normal plots of the raw differences, $D(\mathbf{x},t)$, and the adjusted residuals at one of the monitoring sites. The raw differences have far fatter tails than a normal distribution. The adjusted residuals are much closer to normal, but still with some clear deviations from normality in the extreme tails, especially the upper tail. Quadrupling the number of knots to 288 in the spline fit to the volatility does not change this plot substantially. Thus, although the devolitalization procedure helps greatly in making the process closer to Gaussian, it does not completely solve the problem.

Let us now return to the issue of differencing the observations at each site. In addition to the problems with interpretability noted earlier, the other disadvantage of differencing is that we will have to somehow "undifference" our predictions at unmonitored locations to get actual pressure predictions. Both of these difficulties could be avoided by analyzing the undifferenced pressure data. However, spectral analysis would become highly problematic due to the enormous dynamic range of the undifferenced pressure spectra. More fundamentally, it is not clear to me how one would remove the changes in volatility without first using some kind of high pass filter, of which differencing is an example. In particular, the spatial variation at a given time of the elevation-adjusted pressure values does not even remotely track the changes in variability shown in Figure 2. As an alternative to differencing, one could, at each site, compute the residuals from a running moving average of sufficiently short duration and divide these residuals by an estimated volatility to make the process more nearly Gaussian, but then one would have to model the moving average process and its residuals as a bivariate spatial-temporal process in order to produce prediction intervals at an unmonitored sites. This appears to me to be rather more challenging than what I will do in Section 4, in which I just have to model the spatial pattern of average pressure over the month in order to convert predictions of first differences into predictions of pressure.

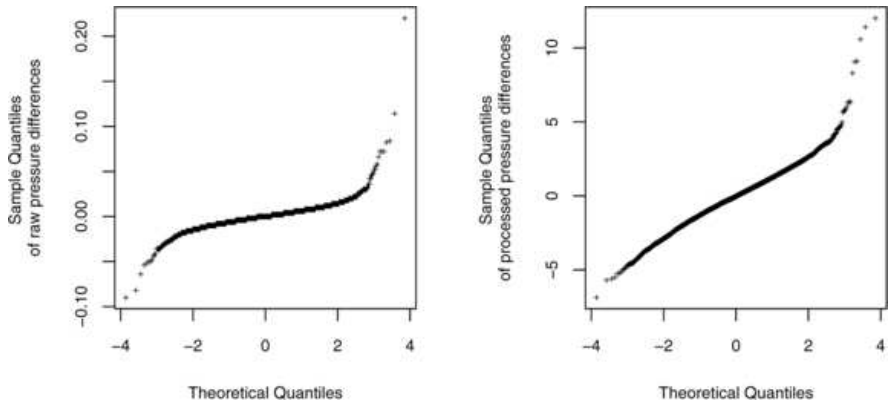

FIG. 3. *Normal plot of $D(\mathbf{x},t)$, the first differences of the pressure (left) and of $A(\mathbf{x},t)$, the adjusted residuals (right) at the site in the northwest corner of Figure 1.*

**3. Model.** Let $K(\mathbf{x},t)$ be the spatial-temporal autocovariance function for the process $A(\mathbf{x},t)$. For data taken regularly in time at a modest number of sites, Stein (2005) argued that the following representation for $K$ is helpful for modeling and estimation:

$$K(\mathbf{x},t) = 1\{\mathbf{x} = \mathbf{0}\} \int_{-\pi}^{\pi} S_0(\omega) e^{i\omega t}\, d\omega$$
(1)

$$+ \int_{-\pi}^{\pi} S_1(\omega) C(|\mathbf{x}|\gamma(\omega)) e^{i\mathbf{u}'\mathbf{x}\theta(\omega)+i\omega t} \, d\omega,$$

with $1\{\cdot\}$ an indicator function, $S_0$ and $S_1$ even integrable functions, $C$ an isotropic correlation function on $\mathbb{R}^2$, $\gamma$ an even positive function, $\theta$ an odd function and $\mathbf{u}$ a unit vector. Stein (2005) also considered a version of this model in which the spatial domain is the surface of a sphere, but the observation domain is fairly small here and I will act as if it is flat (although I compute distances between sites using the great circle distances). The function $S = S_0 + S_1$ gives the marginal spectral density of the process at any site. The decomposition of $S$ into two terms, $S_0$ and $S_1$, appears to improve the fit substantially. Gneiting (2002) calls the contribution of $S_0$ to $K$ the spatial nugget. The function $C$ gives the spatial correlation function of $A$ at any given time, $\gamma$ determines (along with $C$) the coherence between time series at different locations and $\theta$ and $\mathbf{u}$ the phase relationships. See Stein (2005) for further details.

I will need several critical modifications of the approach used in Stein (2005) to fit the pressure data here. Stein (2005) used series expansions for the functions $S_0, S_1, \gamma$ and $\theta$, specifically, cosine functions for the logarithms of the even nonnegative functions $\gamma, S_0$ and $S_1$ and sine functions for the odd $\theta$. This approach worked well enough for daily wind data, but is rather poorly suited for the high frequency data here in which most of the variation in the functions is concentrated in the lower temporal frequencies and the coherences are negligible at higher frequencies. For example, Figure 4 plots empirical coherences (actually, the modulus of the complex coherence times the sign of its real part) based on lightly smoothed periodograms and cross-periodograms for the two nearest and two most distant pairs of sites up to the 50-minute frequency. Not surprisingly, the estimated coherences are stronger for the nearest pair of sites and at lower frequencies, but even at the nearest pair of sites, the plot shows no sign of coherence at frequencies beyond the hourly. To capture these patterns, I will use cubic B-splines as basis functions and then place a higher concentration of knots at the lower frequencies to reflect the expectation that $S_0, S_1$ and especially $\gamma$ and $\theta$ have greater variation at these frequencies.

The lack of coherence at higher frequencies in Figure 4 indicates that $\gamma$ should be very large at these frequencies, which leads to unstable parameter estimates when using B-splines or other localized basis functions. Thus, I replace $\gamma$ by $\delta(\omega) = 1/\gamma(\omega)$, so that $\delta(\omega)$ should be near 0 at higher frequencies. Indeed, based on Figure 4 and other evidence, I set $\delta(\omega) = 0$ for $|\omega| > \omega_0$, where $\omega_0$ is the hourly frequency and the coherence is set to 0 whenever $d > 0$ and $\delta(\omega) = 0$. To make $\delta$ a smooth function of $\omega$ for all $\omega$ (including $\omega_0$), on $(-\omega_0, \omega_0)$, I take $\delta$ as a linear combination of B-spline basis functions with knots at $0, \pm\omega_{\delta 1}, \ldots, \pm\omega_{\delta b}$, where $0 < \omega_{\delta 1} < \cdots < \omega_{\delta b} = \omega_0$, and constrain the coefficients of the B-spline to make $\delta$ even and $\delta(\omega_0) = \delta'(\omega_0) = \delta''(\omega_0) = 0$, yielding $b-1$ independent parameters. Without additional constraints on these coefficients, the resulting $\delta$ may not be nonnegative. Requiring the coefficients to be all nonnegative is sufficient but not necessary to make $\delta$ nonnegative. Rather than enforcing such a constraint, which may lead to parameter estimates falling on a boundary of the parameter space, I replace $C(|\mathbf{x}|\gamma(\omega))$ in (1) by $C(|\mathbf{x}|/|\delta(\omega)|)$. Because the correlation function $C$ used here has the property that the function and all of its derivatives decay exponentially at large arguments, $C(|\mathbf{x}|/|\delta(\omega)|)$ is, in fact, infinitely differentiable in $\delta(\omega)$ at $\delta(\omega) = 0$. Of course, $\delta$ and $-\delta$ yield the same $K$, so there is a trivial lack of identifiability in the model.

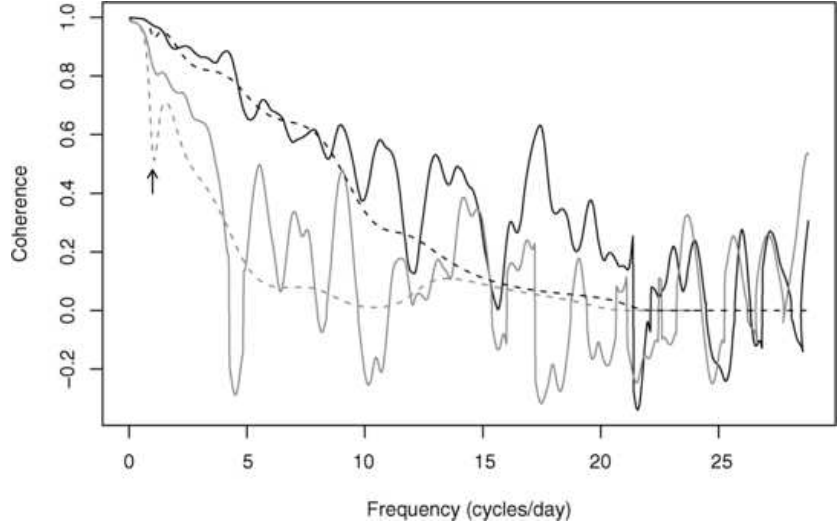

FIG. 4. *Empirical (solid lines) and fitted (dashed lines) coherences up to the 50-minute frequency for the two nearest (black) and two most distant (gray) sites. The arrow highlights the daily frequency.*

For the function $\theta$ controlling the phase relationships, I place knots at $0, \pm\omega_{\theta 1}, \ldots, \pm\omega_{\theta c}$, where $0 < \omega_{\theta 1} < \cdots < \omega_{\theta c} = \omega_0$, and constrain the coefficients of the B-spline to make $\theta$ odd and $\theta(\omega_0) = \theta'(\omega_0) = \theta''(\omega_0) = 0$, yielding $c-1$ independent parameters. I set $\theta(\omega) = 0$ for $|\omega| > \omega_0$, although there is some redundancy here as the phase relationship is irrelevant at frequencies for which the coherence is 0. Rather than fixing the direction $\mathbf{u}$ to be from the west as in Stein (2005), I allow $\mathbf{u}$ to be estimated. Note, then, that $(\theta, \mathbf{u})$ and $(-\theta, -\mathbf{u})$ correspond to the same model.

Next, consider the models for $S_0$ and $S_1$. For $|\omega| > \omega_0$, there is no need to distinguish between these terms, since the coherence is 0 at these frequencies, whichever function is allocated power. To obtain smoothness in $S$ at all frequencies, including $\omega_0$, instead of modeling $S_0$ and $S_1$, I model $S$ and $\beta(\omega) = \log\{S_1(\omega)/S_0(\omega)\}$. More specifically, since the value of $\beta$ is irrelevant for $|\omega| > \omega_0$, I model $\beta$ on $(-\omega_0, \omega_0)$ using a constant function and B-splines with knots at $0, \pm\omega_{\beta 1}, \ldots, \pm\omega_{\beta d}$, where $0 < \omega_{\beta 1} < \cdots < \omega_{\beta d} = \omega_0$, and constrain the coefficients of the B-spline to make $\beta$ even and $\beta'(\omega_0) = \beta''(\omega_0) = 0$ [but not $\beta(\omega_0) = 0$], yielding $d$ independent parameters. To model $S$, I use a B-spline basis with knots at $0, \pm\omega_{S1}, \ldots, \pm\omega_{Se}$, where $0 < \omega_{S1} < \cdots < \omega_{Se} = \pi$ and the coefficients of the B-spline constrained to make $S$ even with $S'(\pi) = 0$, yielding $e+1$ independent parameters. Note that because $\delta(\omega)$ is 0 for $|\omega| > \omega_0$ and is twice differentiable at $\omega_0$, the coherence and phase spectra between any two sites are also twice differentiable at $\omega_0$ despite the fact that $\beta$ is not constrained to be continuous at $\omega_0$.

Finally, for the isotropic correlation function $C$, I use $C(r) = e^{-r}(1+r)$, a Matérn correlation function with smoothness parameter $\frac{3}{2}$, which corresponds to a process that is exactly once mean square differentiable in any direction [Stein (1999)]. The thinking behind this choice is that pressure fields ought to be fairly smooth; choosing correlation functions for yet smoother processes did not lead to improved fits. There is no need to include a separate range parameter in $C$, since multiplying $\delta$ by a scalar factor is identical to changing the range by that factor.

**4. Results.** For fixed knot locations of the functions $S, \delta, \beta$ and $\theta$, one can then, using the multivariate Whittle likelihood, easily approximate the likelihood function in the spectral domain [see Stein (2005)] based on the usual approximations that the (multivariate) periodogram is independent at distinct Fourier frequencies and the expected values of this periodogram are given by the (matrix-valued) spectral density. I included the zero frequency in my likelihood approximation since the mean of $A(\mathbf{x}, t)$ should be effectively 0 and I will need a value for the spectral density at this frequency to generate my predictions. The Whittle approximation is improved by having differenced the process, which greatly reduces the dynamic range of the marginal spectra. The constraint in the model that there is no coherence for $|\omega| > \omega_0$ further speeds the computations, since the covariance matrix of the multivariate periodogram at these frequencies is then just a multiple of the identity matrix. For given knot locations, I estimated the parameters by maximizing the Whittle likelihood using the nlm routine in R.

Choosing the numbers and locations of the knots was done by "hand," iteratively adding, deleting and moving knots until I found what I felt was a good compromise between goodness-of-fit (as measured by the maximized Whittle likelihood) and parsimony. The total number of parameters in the final fit was 29: 1 for the direction of $\mathbf{u}$, 9 for $S$, 4 for $\beta$, 3 for $\theta$ and 12 for $\gamma$. My strategy was to keep the knots fairly regularly spaced but with a tendency to have more knots at the lower frequencies. The actual locations for the knots are given in the appendix. In the final fitted model, one of the coefficients for $\delta$ turned out to be slightly negative, but the estimated $\delta$ itself was everywhere nonnegative.

It is apparent that regular use of this modeling approach would require a more automated approach to knot selection. There is a substantial literature on automatic knot selection for regression splines [see, e.g., Biller and Fahrmeir (2001); Friedman (1991); Lee (2000, 2002); Leitenstorfer and Tutz (2007); Molinari, Durand and Sabatier (2004); Osborne, Presnell and Turlach (1998); Zhou and Shen (2001)], although these works do not explicitly address spectral estimation. Pawitan and O'Sullivan (1994) used smoothing splines to estimate the spectrum of a univariate time series and Pawitan (1996) of a bivariate time series. Dai and Guo (2004) and Rosen and Stoffer (2007) extended these methods to multivariate spectra. None of these works on spectrum estimation allow variable amounts of smoothing across frequencies; nor are they directly applicable to the present setting in which the multivariate spectrum [which includes $\binom{n+1}{2}$ distinct spectra and cross-spectra] is modeled in terms of just 4 functions of frequency. Nevertheless, it should, in principle, be possible to adapt their approaches to the present setting and to allow variable amounts of smoothing across frequencies by including a weight function in the smoothness penalty.

Figure 4 shows that the fitted model does a good job of tracking the empirical coherences. However, there are some signs of misfit, including the strange dip in the fitted coherences around the daily frequency and some underestimation of the coherence for the two nearest sites for frequencies between around 10 and 20 cycles per day. Figure 5 shows the averages over the 11 sites of the unsmoothed periodograms at the 11 sites and the fitted marginal spectrum, with frequency plotted on the log scale to highlight the lower frequencies. There is perhaps some evidence of misfit at the lowest frequencies, although one has to keep in mind that, due to the strong coherence at these frequencies, the corresponding periodogram values at the 11 sites are strongly correlated, so that the empirical spectrum at these frequencies is highly variable.

The main goal in this work is to predict pressure at the two sites left out of this analysis. Let us first consider predicting the first differences of the five-minute average pressure, $\Delta(\mathbf{x}, t) = Z(\mathbf{x}, t+1) - Z(\mathbf{x}, t)$, which differs from $D(\mathbf{x}, t)$ in that these pressure differences have not been corrected to sea level. Specifically, I generated 99 conditional simulations of $\Delta(\mathbf{x}, t)$ at the two sites for $t = 1, \ldots, 8640$. To take some account of the uncertainty in the parameter estimates, instead of using the maximum likelihood estimates in each simulation, I simulated 99 sets of parameter values from the multivariate normal distribution with mean given by the maximum likelihood estimates and covariance matrix by the inverse Hessian of the loglikelihood evaluated at its maximum. The simulations of $A(\mathbf{x}, t)$ were carried out in the spectral domain. Specifically, defining $\widehat{A}(\mathbf{x}, \omega) = \sum_{t=1}^{8640} A(\mathbf{x}, t)e^{i\omega t}$, at each

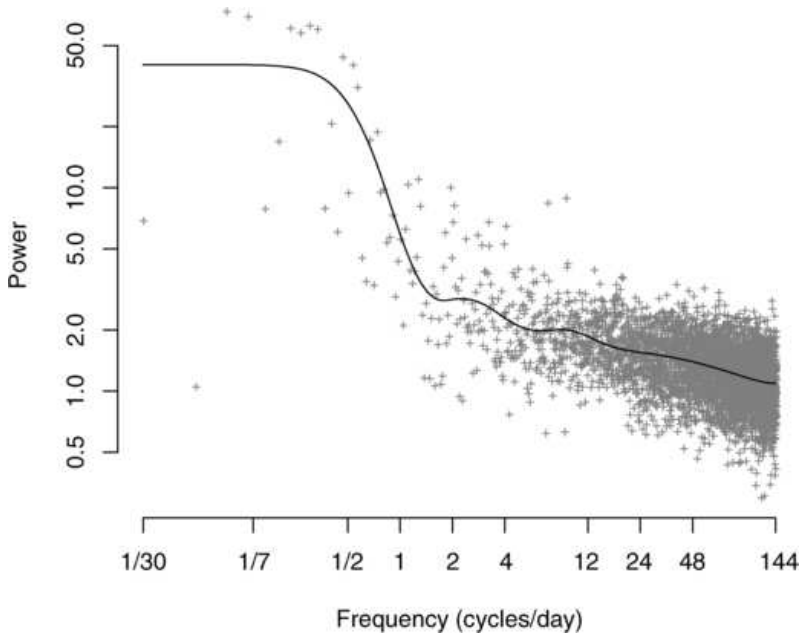

FIG. 5. *Average (over 11 sites) of raw marginal periodograms (+'s) and fitted marginal spectrum (solid curve).*

Fourier frequency, I independently simulated $\widehat{A}(\mathbf{x},\omega)$ from the appropriate conditional bivariate (complex) normal distribution, then recovered the simulated $A(\mathbf{x},t)$ values by taking the inverse discrete Fourier transform. Note that, for $|\omega| > \omega_0$, $\widehat{A}(\mathbf{x},\omega)$ at the prediction sites is independent of $\widehat{A}(\mathbf{x},\omega)$ at the observed sites, speeding the simulations. If, instead of varying the parameter values across simulations, the maximum likelihood estimates are used in each of the 99 simulations, then the mean over the 8640 time points of the sample variances of the 99 simulated values of $A(\mathbf{x},t)$ is lessened by only about a quarter percent at each of the two prediction sites, so perhaps accounting for uncertainty in the number and location of the knots would not matter much either.

The simulated $A(\mathbf{x},t)$ processes obtained in this way have a period of 30 days. Thus, this approach is not appropriate for predicting future pressure. However, to interpolate in space at times at which we have observations, the periodicity of the simulated $A(\mathbf{x},t)$ process may have only a modest effect on the simulations of the undifferenced pressure and then mainly at the very beginning and end of the time period. The inclusion of the 0 frequency in the conditional simulations of $A(\mathbf{x},t)$ prevents $\sum_{t=1}^{8640} A(\mathbf{x},t)$ from equaling 0, thus avoiding one possible problem with this approach. The estimated multivariate spectral density is available at all frequencies in $(-\pi, \pi)$, and not just the Fourier frequencies, so it is possible to calculate the actual estimated $K(\mathbf{x}, t)$ as accurately as desired by numerically integrating (1) over a dense grid of $\omega$ values. One could then conditionally simulate $A(\mathbf{x}, t)$ directly in the space–time domain, although the computational burden would be much heavier than here, where I have done independent simulations at every frequency. Furthermore, given that the likelihood was approximated assuming independence in the frequency domain, it is not clear that a conditional simulation that avoided this assumption would actually be better than the simulations used here when predicting at observed time points.

The simulated $\Delta(\mathbf{x}, t)$ series are obtained by multiplying the simulated $A(\mathbf{x}, t)$ series by $V(t)$, adding back in the diurnal cycle, adjusting the pressure to the appropriate altitudes by inverting the relationship used to correct to sea level, then differencing. Figure 6 shows that the overall coverage properties of the simulated series are very good at the peripheral prediction site and modestly overconservative at the central site. This overconservatism may be due to the underestimation of the coherence at middle frequencies noted in Figure 4. Figure 7 shows a similar plot for first differences in the hourly averages of pressure. Taking into account the greater variability due to the lesser number of time points, the results are good at both sites.

These plots only consider the marginal coverage over time. Considering the dramatic changes in variability over time, it is worthwhile to look at coverage properties over a subset of times when $V(t)$ is large. If one selects the times $t$ corresponding to the largest 10% of $V(t)$ values, the coverage is fairly good, but Figure 8 shows that the observed differences are too often the largest or smallest among the simulated values and too infrequently of ranks between 2 and 20 or 80 and 99. Figure 9 shows the observed and the first two simulated series of pressure differences over October 18–20, which includes the period of greatest volatility. We see that the simulated curves mimic the magnitude of local variations reasonably well during the periods of lower volatility, but, during the period of high volatility, the magnitudes of the largest simulated pressure differences are not sufficiently large compared to the observed pressure differences. Using a larger number of degrees of freedom in the smoothing spline may have helped somewhat here, but I think a better solution is to take a completely different approach to estimating the volatility (see Discussion).

Overall, the inferences for $\Delta(\mathbf{x}, t)$ at the prediction sites are quite well calibrated. However, to simulate the pressure rather than its differences requires inference about some linear combination of $Z(\mathbf{x}, t)$ values that is not a function of the $\Delta(\mathbf{x}, t)$'s, such as $Z(\mathbf{x}, 1)$. Instead of trying to model the joint distribution of $Z(\mathbf{x}, t)$ at any one time and the $\Delta$ process, I will conditionally simulate $\bar{Z}(\mathbf{x}) = \frac{1}{8640} \sum_{t=1}^{8640} Z(\mathbf{x}, t)$ at the two prediction sites and assume $\bar{Z}$ is independent of $\Delta$. Because I only have $\bar{Z}$ at 11 sites, I need to use a very simple model for this spatial process. Specifically, writing $a(\mathbf{x})$ for the altitude at location $\mathbf{x}$, I take $M(\mathbf{x}) = \bar{Z}(\mathbf{x})\exp\{a(\mathbf{x})/8310\}$ to be a stationary (or intrinsic) Gaussian process with spatial variogram of the form $\theta G(d)$, with $G$ a valid variogram model and $\theta$ unknown. To estimate $\theta$, I used a restricted maximum likelihood based on the 11 $M$ values available. I simulated $M$ at the two prediction sites from a bivariate $t$ distribution with 10 degrees of freedom to account for the uncertainty in the estimate of $\theta$ [see Handcock and Stein (1993)]. I then undid the corrections to sea level to obtain simulated values of $\bar{Z}(\mathbf{x})$. As noted in Section 2, there appears to be a weak but noticeable spatial pattern to the $M(\mathbf{x})$ values, so I tried taking $G(d) = d$, the linear variogram. However, the loglikelihood of a model with no spatial dependence (a pure nugget effect) is within 0.25 of the linear variogram model, so, in the interest of conservatism, I chose to view the pure nugget model as my "primary" model for prediction, although I also produced predictions using the linear variogram model. Including un-certainty in $M$ substantially increases the variability across simulations. For example, at the central prediction site, the mean across time of the sample variances at each time of the 99 simulations is, relative to simulations with no variation in $M$, 76% larger when using a pure nugget variogram for $M$ and 34% larger when using a linear variogram.

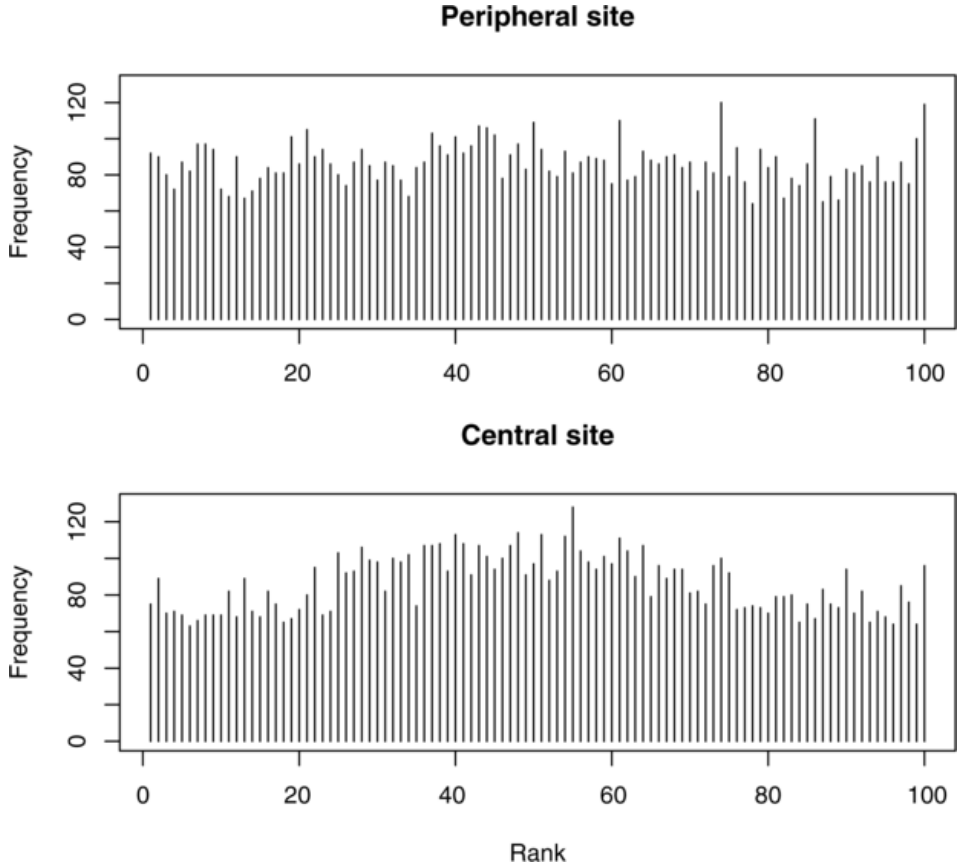

Fig. 6. *Histogram of ranks of observed first differences in pressure among the simulated values over the 8640 time points.*

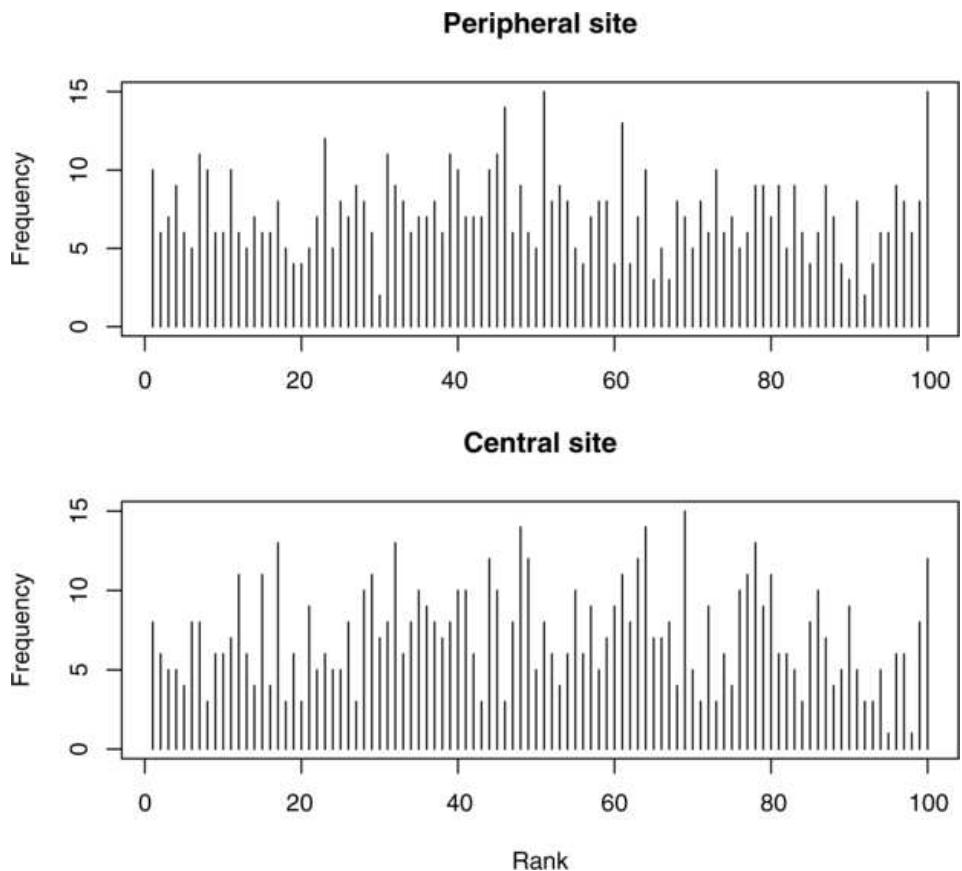

Fig. 7. *Histogram of ranks of observed first differences of hourly pressure among the simulated values over the 719 available hourly differences.*

Figure 10 shows observed pressure and the envelope of the 99 simulated pressure series at the two prediction sites. The simulated pressure tracks the observed pressure quite well; Table 1 provides some summary statistics for the predictor obtained by averaging the 99 series at each time point. The second row of Table 1 gives results when a linear variogram instead of a pure nugget effect is used for $M$. The last row of Table 1 shows results for a simple nearest neighbor predictor: predict $Z(\mathbf{x},t)$ by $Z(\mathbf{x}',t)$ (adjusted for elevation), where $\mathbf{x}'$ is the location of the monitoring site closest to $\mathbf{x}$. At the peripheral site, the standard deviation of the errors for the nearest

**Peripheral site**

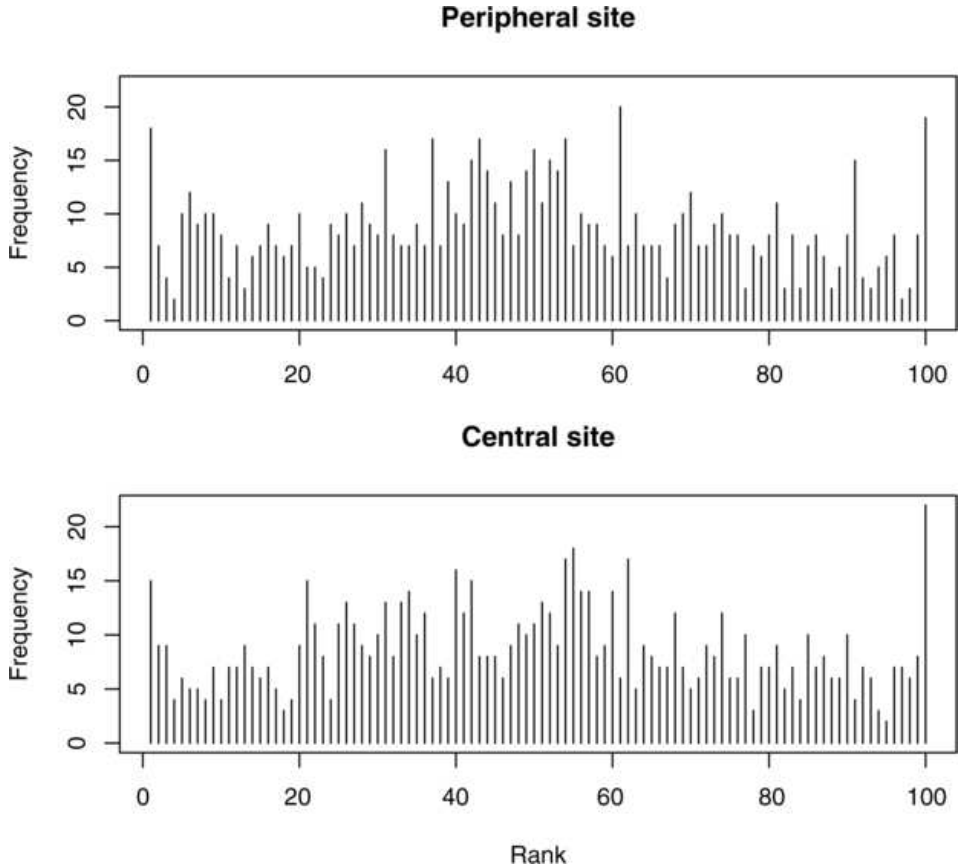

**Central site**

Fig. 8. *Histogram of ranks of observed first differences of hourly pressure among the simulated values over the 10% of times with the highest volatilities.*

neighbor predictor is more than 3 times as large as the other predictors and is about 50% larger at the central site. Table 1 shows that the average error is generally a substantial component of the root mean squared error. However, the fact that one method might have smaller average errors at one site or the other is not very informative since this advantage could easily be due to luck. We can conclude that if it were possible to come up with better predictions of $\bar{Z}(\mathbf{x})$, these could lower the root mean squared prediction errors substantially.

The simulation envelopes at both prediction sites are shown in Figure 10. Here, I will focus on the peripheral site; qualitatively similar results hold at the central site. The mean width of the simulation envelope at the peripheral site is 0.34 kPa (the mean width of the 90% prediction intervals is 0.21 kPa). Despite this rather narrow width, the simulation envelope is far too conservative, with the observed pressure being outside the envelope only 17 times as opposed to the expected value of 168 if the intervals were calibrated. Furthermore, the upper envelope is particularly conservative, as the observed pressure is one of the highest 30 ranks only 0.56% of the time as opposed to 30% of the time for a calibrated interval. This asymmetry in the upper and lower envelopes should be expected, given the fact that the predicted values for $\bar{Z}(\mathbf{x})$ are higher on average than the observed $\bar{Z}(\mathbf{x})$. Since the prediction errors at different times may be strongly correlated, this apparent substantial miscalibration is not necessarily a sign of a problem with the methodology. For example, if we consider the ranks at each time point of the 100 series (the observed series plus the 99 simulated series), then 19 of these series are at no time the minimum or maximum of the 100 series. Thus, being outside the simulation envelope only 17 times is not at all unusual. This overconservativeness is not caused by too much variability in the simulated values of $\bar{Z}$, since if we set $\bar{Z}$ to the same value in all 99 simulated series (given by the elevation-adjusted kriging predictor under the pure nugget model), then the observed series is the minimum or maximum at 10 time points, whereas 14 of the 99 simulated series are never the minimum or maximum.

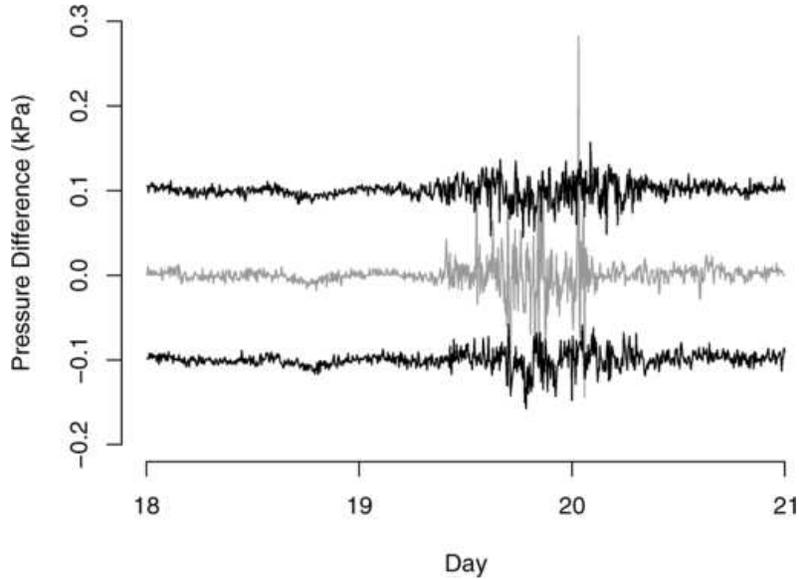

Fig. 9. *Observed (gray curve) and simulated (two black curves) pressure differences at peripheral site from October 18–20. Simulations are offset by $\pm 0.1$ kPa for legibility.*

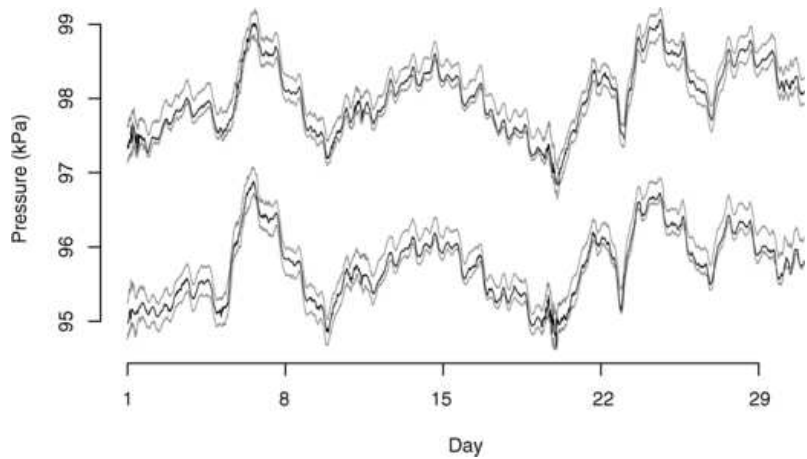

Fig. 10. *Observed (black curve) and pointwise maxima and minima (gray curves) of the 99 simulations at the central (upper curves) and peripheral site (lower curves).*

Table 1
*Sample means and standard deviations of prediction errors and root mean squared errors at the two prediction sites. The first two rows use the main model presented here, the only difference being whether a pure nugget effect or a linear variogram is used to model the spatial variations in $M(\mathbf{x})$. The last row gives results for an elevation-adjusted nearest neighbor predictor*

| Model | Peripheral site | | | Central site | | |
|---|---|---|---|---|---|---|
|  | Mean | st dev | rmse | Mean | st dev | rmse |
| Pure nugget | −0.048 | 0.029 | 0.056 | −0.035 | 0.025 | 0.044 |
| Linear | −0.050 | 0.029 | 0.058 | −0.014 | 0.025 | 0.029 |
| Nearest neighbor | −0.048 | 0.091 | 0.103 | −0.010 | 0.037 | 0.039 |

**5. Discussion.** This work only considers prediction of a single meteorological quantity at two locations. Extending the approach to a large number of locations and/or to predicting area averages introduces no new conceptual challenges, although the computational burden would increase. However, the assumption of no coherence at temporal frequencies higher than hourly is unreasonable at sufficiently small spatial scales, so that conditional simulations at high spatial resolution under the fitted model here would have high frequency fluctuations with too much spatial variability at nearby locations. To examine the scope of this problem, I carried out conditional simulations at two locations, the peripheral prediction site and a location 1 km north of this site (and assumed to be at the same elevation). Over the 99 simulations, the average variance of the differences between the simulated values at the two sites was 0.0030, which is about 14% of the same quantity when comparing the simulations at the peripheral and central prediction sites. This average variance corresponds to a standard deviation of 0.054 kPa, which is not that much bigger than the overall uncertainty in the measurements of 0.035 kPa.

Nevertheless, if one wanted to do something about even these quite small local fluctuations in pressure, one could change the model to allow spatial dependencies at higher frequencies. For example, consider setting $\omega_0 = \pi$ (so that $\delta$ is only forced to be 0 at $\pm\pi$), thus adding a knot at $\pi$ for $\delta$, $\theta$ and $\beta$, but otherwise leaving all of the other knot locations given in the appendix unchanged. The lack of spatial independence at higher frequencies does slow down the computations relative to the model with $\omega_0 = \pi/6$, but they are still manageable. The loglikelihood then increases by about 73 with the addition of the three parameters, even though the fitted coherences at the hourly frequency, which were forced to be 0 in the smaller model, have quite small values: between 0.0079 and 0.068 for all pairs of monitoring sites. Unfortunately, making this change does not, in fact, create strong coherences at high frequencies and small spatial scales because the estimated value for $\beta$ is less than $-1.88$ for all frequencies higher than the hourly, putting an upper bound of around 0.132 on coherences in this frequency range no matter how close two locations are. Thus, it would appear that we would need to remove or somehow constrain the spatial nugget effect at higher frequencies in order to get the strong coherences we would want at short spatial scales. Removing the spatial nugget entirely seriously degrades the fit at lower frequencies. Replacing the spatial nugget term in the first line of (1) by, for example, $\int_{-\pi}^{\pi} S_0(\omega) C_0(|\mathbf{x}|) e^{i\omega t} \, d\omega$, where $C_0$ is a valid, continuous, isotropic correlation function that is identically 0 at all distances greater than the shortest distance between the 11 monitoring sites, would not have any effect on the likelihood function but would allow coherences to tend to 1 as distances tend to 0. However, the data provide no information about the choice of $C_0$, so such a solution would be highly arbitrary. A better solution would be to collect data for some period of time at a small but tightly spaced set of sites as part of one of the SGP Field Campaigns that are carried out in the region (see www.arm.gov/sites/sgp.stm).

Extending this work to the multivariate setting is a greater challenge. Specifically, it is not obvious how to extend the model in Stein (2005) to the multivariate setting in a way that allows for realistic dependencies across quantity, space and time. Although some recent works such as Haas (2002) and Tzala and Best (2008) consider statistical modeling of multivariate space–time processes, these works focus on much longer time scales and it is not clear the models and methods they propose are suitable for capturing the dynamics affecting high-frequency meteorological data. Another challenge in statistically modeling winds and temperatures is that there are clear diurnal cycles in the dependence structure that may not be removable by such simple schemes as rescaling the data depending on the hour of the day. Therefore, it may be necessary to use space–time multivariate models that are only cyclostationary in time [Hurd and Miamee (2007)] rather than stationary.

Finally, let us return to the issue raised in the Introduction of making use of further meteorological information. In particular, such information might be of considerable value in handling the bursts of high volatility. Specifically, to the extent that rapid changes in pressure are due to the passage of weather fronts and the space–time evolution of these fronts can be mapped using, for example, upper level winds, one could try to model the volatility at a particular place and time in terms of a distance to the nearest front and the strength of that front rather than, as I did here, assuming the volatility does not depend on spatial location.

## APPENDIX: KNOT LOCATIONS

For each of the functions $S, \beta, \delta$ and $\theta$, I only considered Fourier frequencies for the knot locations, that is, frequencies of the form $\pi j/4320$ for integer $j$. For all four functions, whenever $\pi j/4320$ is a knot, so is $-\pi j/4320$. The $j$ values for the final knot locations for $S$ are $0, 10, 30, 60, 120, 400, 720, 4320$; for $\delta$, $0, 5, 10, 15, 25, 40, 60, 90, 150, 240, 360, 480, 600, 720$; and for both $\beta$ and $\theta$, $0, 40, 120, 360, 720$.

## REFERENCES

BILLER, C. and FAHRMEIR, L. (2001). Bayesian varying-coefficient models using adaptive regression splines. *Statist. Modell.* **1** 195–211.
DAI, M. and GUO, W. (2004). Multivariate spectral analysis using Cholesky decomposition. *Biometrika* **91** 629–643. MR2090627
FRIEDMAN, J. H. (1991). Multivariate adaptive regression splines (with discussion). *Ann. Statist.* **19** 1–141. MR1091842
GNEITING, T. (2002). Nonseparable, stationary covariance functions for space–time data. *J. Amer. Statist. Assoc.* **97** 590–600. MR1941475
HAAS, T. C. (2002). New systems for modeling, estimating, and predicting a multivariate spatio–temporal process. *Environmetrics* **13** 311–332.
HANDCOCK, M. S. and STEIN, M. L. (1993). A Bayesian analysis of kriging. *Technometrics* **35** 403–410.
HURD, H. L. and MIAMEE, A. (2007). *Periodically Correlated Random Sequences: Spectral Theory and Practice.* Wiley, New York. MR2348769
LEE, T. C. M. (2000). Regression spline smoothing using the minimum description length principle. *Statist. Probab. Lett.* **48** 71–82. MR1767610
LEE, T. C. M. (2002). On algorithms for ordinary least squares regression spline fitting: A comparative study. *J. Statist. Comput. Simul.* **72** 647–663. MR1930486
LEITENSTORFER, F. and TUTZ, G. (2007). Knot selection by boosting techniques. *Comput. Statist. Data Anal.* **51** 4605–4621. MR2364468
MASS, C. F., STEENBURGH, W. J. and SCHULTZ, D. M. (1991). Diurnal surface-pressure variations over the continental United States and the influence of sea level reduction. *Monthly Weather Review* **119** 2814–2830.
MOLINARI, N., DURAND, J.-F. and SABATIER, R. (2004). Bounded optimal knots for regression splines. *Comput. Statist. Data Anal.* **45** 159–178. MR2045466

OSBORNE, M. R., PRESNELL, B. and TURLACH, B. A. (1998). Knot selection for regression splines via the lasso. *Comput. Sci. Statist.* **30** 44–49.

PAWITAN, Y. and O'SULLIVAN, F. (1994). Nonparametric spectral density estimation using penalized Whittle likelihood. *J. Amer. Statist. Assoc.* **89** 600–610. MR1294086

PAWITAN, Y. (1996). Automatic estimation of the cross-spectrum of a bivariate time series. *Biometrika* **83** 419–432. MR1439793

RITSCHE, M. T. (2008). Surface meteorological observation system handbook. DOE SC-ARM TR-031, Atmospheric Radiation Measurement Program, U.S. Department of Energy. Available at www.arm.gov/publications/tech_reports/handbooks/smos_handbook.pdf.

ROSEN, D. and STOFFER, D. S. (2007). Automatic estimation of multivariate spectra via smoothing splines. *Biometrika* **94** 335–345. MR2331489

RUBIN, D. B. (1987). *Multiple Imputation for Nonresponse in Surveys*. Wiley, New York. MR0899519

SCHNEIDER, T. (2006). Analysis of incomplete data: Readings from the statistics literature. *Bull. Amer. Meteor. Soc.* **87** 1410–1411.

STEIN, M. L. (1999). *Interpolation of Spatial Data: Some Theory for Kriging*. Springer, New York. MR1697409

STEIN, M. L. (2005). Statistical methods for regular monitoring data. *J. Roy. Statist. Soc. Ser. B* **67** 667–687. MR2210686

TZALA, T. and BEST, N. (2008). Bayesian latent variable modelling of multivariate spatio-temporal variation in cancer mortality. *Statist. Methods Medical Research* **17** 97–118.

ZHOU, S. and SHEN, X. (2001). Spatially adaptive regression splines and accurate knot selection schemes. *J. Amer. Statist. Assoc.* **96** 247–259. MR1952735

DEPARTMENT OF STATISTICS
UNIVERSITY OF CHICAGO
CHICAGO, ILLINOIS 60637
USA
E-MAIL: stein@galton.uchicago.edu